\documentclass[letterpaper, 10 pt, conference]{support/ieeeconf}
\usepackage{times}
\usepackage[pdftex]{graphicx}
\usepackage{subfigure}
\usepackage{amsmath,amssymb,amsopn,amstext,amsfonts}
\usepackage{cancel}
\usepackage[space]{cite}
\usepackage{pdfsync}
\usepackage{balance}
\usepackage{color}
\usepackage{mathtools}
\usepackage{algorithm}
\usepackage{algorithmic}
\usepackage{bm}

\usepackage{diagbox}
\usepackage{float}
\usepackage{epstopdf}
\usepackage{url}
\usepackage{multirow}
\usepackage{tikz}
\usepackage[linkcolor=black,citecolor=black,urlcolor=black,colorlinks=true]{hyperref}

\graphicspath{{figures/}}
\DeclareGraphicsExtensions{.pdf,.png,.jpg,.eps}
\IEEEoverridecommandlockouts
\title{\LARGE \bf Full Attitude Control of an Efficient Quadrotor Tail-sitter VTOL UAV with Flexible Modes}
\author{Wei Xu$^{1}$, Haowei Gu$^{2}$, Youming Qing$^{1}$, Jiarong Lin$^{1}$, Fu Zhang$ ^{1} $ 
  \thanks{$^{1}$authors are with the Mechatronics and Robotic Systems (MaRS) Laboratory, Department of Mechanical Engineering, University of Hong Kong, China.
          {\tt\small xuweii, qym96, zivlin, fuzhang\}@hku.hk}}
  \thanks{$^{2}$author is with the Department of Electronic and Com-
puter Engineering, Hong Kong University of Science and Technology,
Hong Kong, China.
            {\tt\small hguad@ust.hk }}
}

\begin{document}

\maketitle
\thispagestyle{empty}
\pagestyle{empty}

\begin{abstract}

In this paper, we present a full attitude control of an efficient quadrotor tail-sitter VTOL UAV with flexible modes. This control system is working in all flight modes without any control surfaces but motor differential thrusts. This paper concentrates on the design of the attitude controller and the altitude controller. For the attitude control, the controller's parameters and filters are optimized based on the frequency response model which is identified from the sweep experiment. As a result, the effect of system flexible modes is easily compensated in frequency-domain by using a notch filter, and the resulting attitude loop shows superior tracking performance and robustness. In the coordinated flight condition, the altitude controller is structured as the feedforward-feedback parallel controller. The feedforward thrust command is calculated based on the current speed and the pitch angle. Tests in hovering, forward accelerating and forward decelerating flights have been conducted to verify the proposed control system.

\end{abstract}

\section{Introduction}
\label{sec:intro}

These years, unmanned aerial vehicles(UAVs) play a significant role in many civil and military fields, such as patrols, surveying, monitoring, and aerial photography~\cite{kuchemann1978aerodynamic}. There are two main types of traditional aerial vehicles: rotor aerial vehicles and fixed-wing aerial vehicles~\cite{frank2007hover}. The rotary aerial vehicles, such as helicopters and multi-rotors, can hover at a stationary point to execute missions like monitoring, but the power efficiency of the rotary aerial vehicle is lower than fixed-wing aerial vehicles and thus achieving a shorter endurance and range. On the other hand, fixed-wing vehicles although have much higher power efficiency but requires a runway to take off and land. The ``sliding takeoff and landing" is possible for small fixed-wing vehicles but will lead to cost and safety problems. As a consequence, a new type of aerial vehicles which can achieve both power efficiency and vertical takeoff and landing attracts more and more research interests\cite{saeed2015review}. Among the various types of VTOL aircrafts configurations, the tail-sitter aircraft is probably the most concise one because there is no need for rotor tilting mechanisms\cite{mccormick1999aerodynamics}. This paper will concentrate on the tail-sitter VTOL platform.

In our group's previous work~\cite{wang2017design} and~\cite{lyu2017design}, we designed and implemented a few prototypes to verify the tail-sitter UAV concepts. Simulation and experiments including the wind tunnel tests~\cite{zhang2017modeling} and outdoor flight tests~\cite{lyu2017hierarchical} were conducted to show that our vehicles can achieve fully autonomous flight including vertical takeoff, transition, level flight and landing. The disturbance observer~\cite{lyu2018disturbance}~\cite{lyu2018h_} and loop-shaping method~\cite{zhou2018frequency} are applied in the hovering controller design to improve the vehicles' hovering capability in cross-wind.

\begin{figure}[h]
    \begin{center}
        {\includegraphics[width=1\columnwidth]{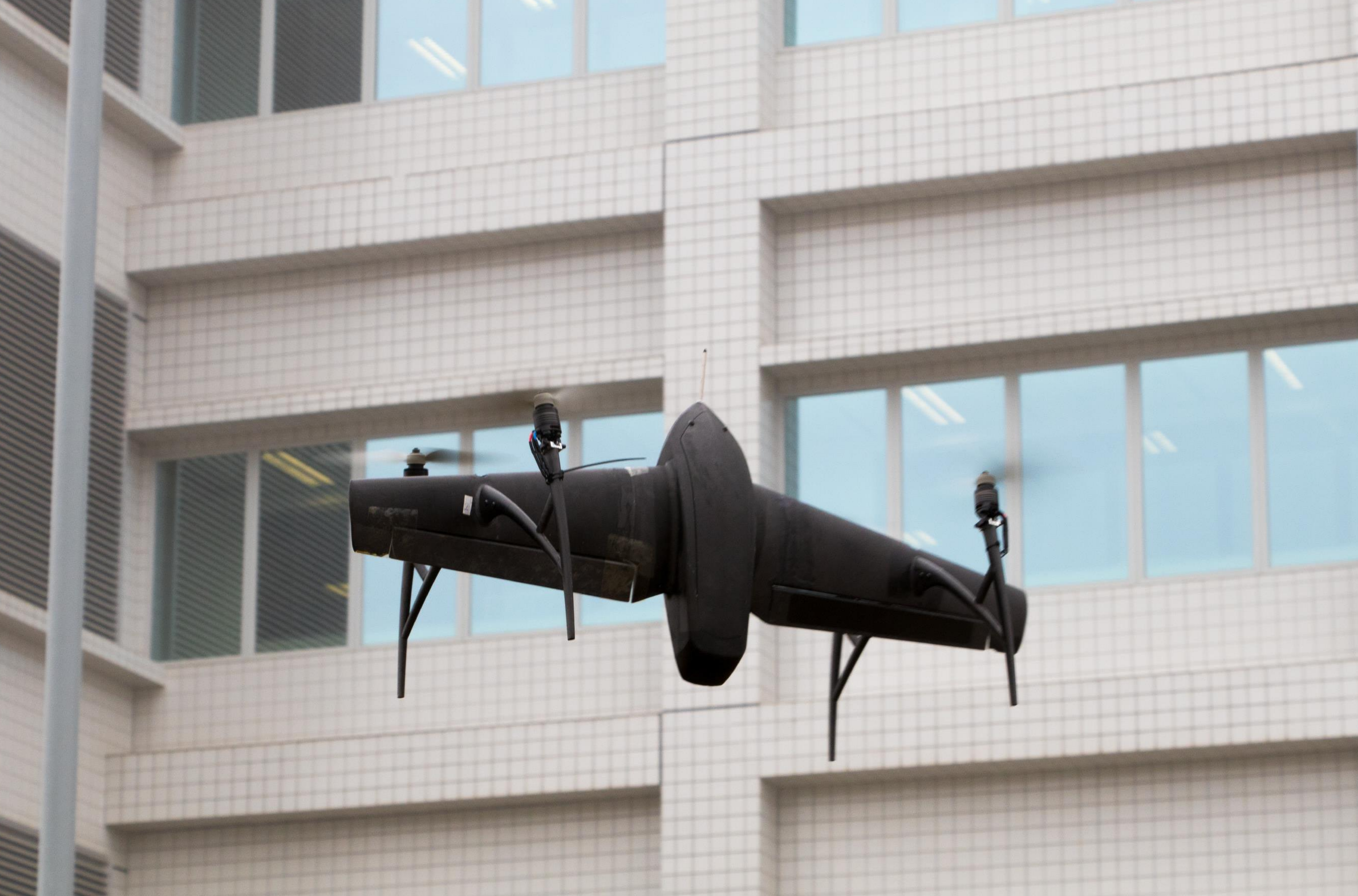}}
    \end{center}
    \vspace{-0.25cm}
    \caption{\label{fig:objective}The quadrotor tail-sitter with carbon-fiber structure}
        \vspace{-0.25cm}
\end{figure}

The previous prototypes were made of foamed plastic and corkwood for fast implementation and verification. In addition, they were largely based on commercial flying wing platforms whose design did necessarily not fit into the tail-sitter configuration in terms of power efficiency. Motivated by this, our recent work~\cite{gu2018coordinate} puts forward a multidisciplinary design and optimization framework to assist this small-scale tail-sitter UAV's design while being subject to various practical constraints~\cite{gu2018coordinate}. The core idea of this work is to formulate the UAV design into a single optimization problem such that the design objective (e.g., power consumption) can be optimized numerically. The optimization problem is solved by a coordinate-descent algorithm, which effectively separates the optimization for propulsion systems from the optimization for UAV airframe. The proposed method leads to an optimal aircraft design whose aerodynamic performance fits extremely well with CFD analysis. Following the optimal aircraft design, a practical carbon-fiber structured tail-sitter UAV with good maneuverability and rigidity was manufactured, which is shown in Fig. \ref{fig:objective}. Actual flight tests show that the power consumption at level flight is five times less than that of in hovering, as predicted by the optimizer as well as CFD analysis.

There are two main differences between the new carbon-fiber structured aircraft and our previous designs in ~\cite{wang2017design}and~\cite{lyu2017design}. The first one is that there is no servo motor or related actuating links, which will lead to a decrease in weight and cost. The control force and torque are only produced by the four rotors, which means a more concise and unified controller can be designed. For the previous configuration, the different mixers are used for the hover, transition, and forward flight~\cite{lyu2017design},~\cite{lyu2018simulation}. In this paper, the unified attitude and altitude controllers are proposed for all the conditions with a large-range angle of attack and velocity. The second difference is that the strength and stiffness of new design are higher, but with some sharp flexible modes possibly due to the articulation imperfection during the assembly process. These sharp flexible modes will limit the bandwidth of angular rates loop. In this paper, a notch filter~\cite{decker1978optical} is designed to decrease the influence of modes.

To design the controller and filter for the new carbon-fiber structured aircraft, an accurate mathematical model of the plant is necessary. The frequency domain model is one popular method to represent the details of the system dynamics such as flexible modes. Pintelon et al.~\cite{pintelon2012system} described the basic concepts of system identification in the frequency domain and proposed some methods to evaluate the precision of identification experiment. Tischler et al.~\cite{tischler2006aircraft} implied the identification method and validation techniques to actual fixed-wing and rotary aircraft systems. Our previous work~\cite{zhou2018frequency} modeled the previous tail-sitter VTOL platform through the discrete sweep experiments in which only a single frequency is used for each sweep experiment. The discrete sweep method can get a precise model at each frequency but desires a large number of the experiments. In this paper, the exponential continuous chirp signal is used to identify the frequency domain model of the carbon-fiber structured tail-sitter VTOL.

The remainder of this paper is organized as follows. Section II will introduce the aircraft platform and its model. The detailed design of the attitude controller and altitude controller will be described in the following section III. Experimental verification is provided in section IV. Finally, section V draws conclusions.

\section{System Configuration}
\label{sec:formulation}

\subsection{Aircraft Design} 
As shown in Fig. \ref{fig:objective}, the quadrotor tail-sitter consists of a carbon-fiber structured airframe and four rotors. The trapezoidal wing with MH-115 airfoil is the main lift producer. The airfoil and structure design optimization can be found in previous works~\cite{gu2018coordinate}. In the latest version, there is no aileron and elevator which means servos are no longer needed. The control moments are produced by the propeller differential thrusts only. This improvement will lead to four benefits. First, servo and relevant structure take up around 5 percent of the total weight, and removing them can reduce the dead weight and increase the load ability. Second, the control surface will lose the control ability when the angle of attack (AOA) is higher than the stall angle while the AOA affects the rotor actuating effect very slightly. This will ease the controller design as well. Third, the servo adds extra dynamics as well as non-linearity arising from the gearbox, therefore elimination of the servo will lead to an increase in control bandwidth. The aircraft configuration parameters are specified in the Table~\ref{table:config}.

\begin{table}[h]
    \newcommand{\tabincell}[2]{\begin{tabular}{@{}#1@{}}#2\end{tabular}}
    \caption{Aircraft Configuration}
    \label{table:config}
    \vspace{-0.15in}
    \begin{center}
        \begin{tabular}{|c|c|}
            \hline
            Propeller name&APC9x6E\\
            \hline
            Motor name&Sunnysky A2212 (980KV)\\
            \hline
            Wing span&0.90 m\\
            \hline
            Taper ratio&0.48\\
            \hline
            Swept angle&7.30${}^\circ $\\
            \hline
            Angle of attack&7.00${}^\circ $\\
            \hline
            Root chord of wing&0.20 m\\
        \end{tabular}
    \end{center}
    \vspace{-0.25cm}
\end{table}

\subsection{Coordinates and Dynamic Model} 
\begin{figure}[h]
    \begin{center}
        {\includegraphics[width=1\columnwidth]{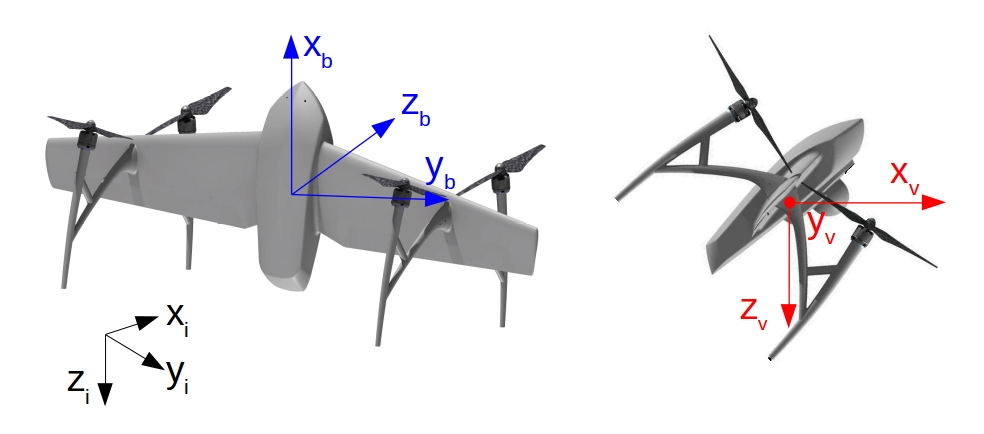}} 
    \end{center}
    \vspace{-0.25cm}
    \caption{\label{fig:coord}The body frame and velocity frame coordinate}
        \vspace{-0.25cm}
\end{figure}

The local north-east-down (NED) coordinate system is chosen as the inertial frame. Body coordinates used in this work are almost as same as those of the conventional fixed-wing aircraft, which is shown in Fig. \ref{fig:coord}. The roll, pitch, and yaw angle are respectively defined as the rotation angle along the x, y, and z-axis of the body coordinate. The order of Tait-Bryan angles is chosen as the Z-X-Y which can avoid the singularity of direct cosine matrix when the pitch angle is near to $ \pi/2 $. The quaternion is also used in the attitude controller to describe the rotation of aircraft. The detail explanation about Tait-Bryan angles, direct cosine matrix, and quaternion can be found in~\cite{zhang2017modeling}. Besides the body frame, the velocity coordinate frame is also defined, where $ x_{v} $ axis points along the ground velocity vector and $ z_{v} $ stay at the plane of symmetry.

The position of the aircraft is denoted by vector $ p^{i}=\left[p_{x}^{i},\,p_{y}^{i},\,p_{z}^{i}\right]^{T} $ which is represented in the inertial frame. The angular rate represented in the body frame is $ \omega_{b} $. The total mass of the aircraft is $ m $, and the inertial matrix is denoted as $ I $. Newton's equations of motion are applied to modeling the rotational and vertical dynamic of aircraft as below:

\begin{equation}
\label{e:plant1}
\begin{array}{l}
 \dot{R_{b}^{i}}= R_{b}^{i}\widehat{\omega}_{b}
 \end{array}
\end{equation}
\vspace{-0.25cm}

\begin{equation}
\label{e:plant2}
\begin{array}{ll}
 I\dot{\omega_{b}}= -\omega_{b} \times\left(I\omega_{b}\right)+\tau+M_{a}
\end{array}
\end{equation}
\vspace{-0.75cm}

\begin{equation}
\label{e:plant12}
\begin{array}{ll}
 \ddot{p}_{z}^{i}=mg+e_{3}\left(f_{a}^{i} + f_{p}^{i}\right)
\end{array}
\end{equation}
\vspace{-0.25cm}

\noindent where $ \widehat{\omega}_{b} $ stands for the skew-symmetric cross product matrix, $ \tau $ is the moment vector produced by the differential motor thrust, and $ M_{a} $ stands for the aerodynamic moment. $ e_{3}= \left[0,\,0,\,1\right] $ is used to separate the $ z^{i} $ axis component from the aerodynamic force $ f_{a}^{i} $ and propeller force $ f_{p}^{i} $. $ R_{b}^{i} $ is the rotation matrix from body frame to inertial frame. The aerodynamic force $ f_{a}^{i} $ and propeller force $ f_{p}^{i} $ can be modeled as:

\begin{equation}
\label{e:aero_2}
 \begin{array}{ll}
 f_{a}^{i} = R_{v}^{i} f_{a}^{v} \quad f_{p}^{i} = R_{b}^{i} f_{p}^{b}
 \end{array}
\end{equation}
\vspace{-0.75cm}

\begin{equation}
\label{e:aero_3}
 \begin{array}{ll}
 f_{a}^{v} = [-D,\, 0,\, -L]^{T} \quad f_{p}^{b} = [T,\, 0,\, 0]^{T}
 \end{array}
\end{equation}
\vspace{-0.5cm}

\noindent where $ R_{v}^{i} $ is the rotation matrix from velocity frame to inertial frame. $ L $ and $ D $ are the aerodynamic lift and drag respectively.  
 
\subsection{Hardwares and Avionics}
The core hardware of our vehicle consists of a Pixhawk 4 Mini controller board, an M8N GPS/GNSS module, and a pitot tube airspeed sensor. Pixhawk4 Mini is currently the lightest and smallest one of the Pixhawk4 serials working with the PX4 open-source flight software stack~\cite{lyu2017hierarchical}. An extended Karman filter is used to estimate the position, linear velocity, angular velocity, and attitude by fusing all the sensor measurements. There are two ground-air communication links. The first one is the 2.4 GHz transmitter and receiver which can pass the remote controller (RC) signals to the controller board. Another one is a two-way 433 MHz telemetry radio which is used to deliver the information between the flying vehicle and a ground station~\cite{autopilotpixhawk}.

\section{Control System Design}

\begin{figure}[h]
    \begin{center}
        \vspace{1.25cm}
        {\includegraphics[width=1\columnwidth]{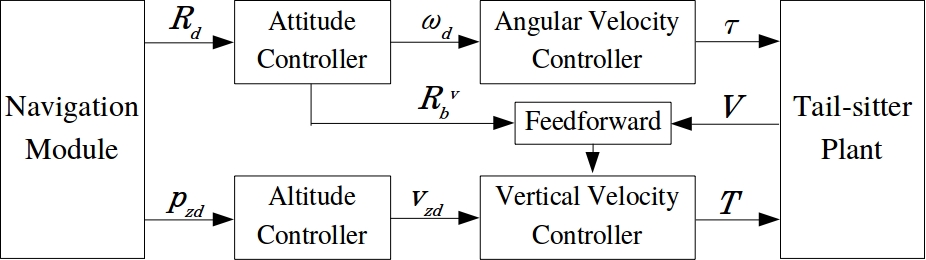}} 
    \end{center}
    \vspace{-0.25cm}
    \caption{\label{fig:ctrl_over}The control structure for Altitude-Attitude flight mode}
        \vspace{-0.5cm}
\end{figure}

The quadrotor tail-sitter VTOL UAV will fly with large flight envelope which means large-range pitch angle and velocity. It is of great significance to make sure that the controller has enough stability and robustness. As shown in Fig. \ref{fig:ctrl_over}, a hierarchical control structure is implied for the Altitude-Attitude flight mode. There are two dual-loop controllers in this structure. The attitude loop and angular velocity loop make up the first controller to track the attitude command. The altitude loop and vertical velocity loop make up the second controller to follow the altitude command. The attitude and altitude commands come from the navigation module which will not be discussed in this paper. The details about the navigation module can be found in~\cite{lyu2017hierarchical} and~\cite{zhou2017unified}.

\subsection{Attitude Controller Design}
\label{sec:attitude_controller}
\begin{figure}[h]
    \begin{center}
        \vspace{0.25cm}
        {\includegraphics[width=1\columnwidth]{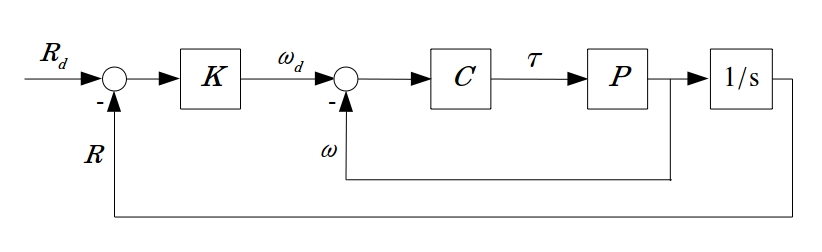}} 
    \end{center}
    \vspace{-0.25cm}
    \caption{\label{fig:atti_and_rate}The attitude and angular velocity controller}
        \vspace{-0.25cm}
\end{figure}

Fig. \ref{fig:atti_and_rate} shows the detailed structure of attitude and angular velocity control loop. The unit negative feedback is used to get the attitude error and angular velocity error. 

$ P $ is the actual aircraft dynamics which is modeled in (\ref{e:plant1}) and (\ref{e:plant2}). $ K $ is the proportional gain from the attitude error to angular velocity command, which is shown as (\ref{e:att_k}). $ \xi_{e} $ is the difference between target attitude and current attitude. when the $ \xi_{e} $ is represented in axis-angle form which is $ \log\left(R_{d}^{T}R\right)$, the result will be singular if the rotation angle is $\pi$~\cite{bullo1995proportional}. Therefore the quaternion ($q_{e}=\left [ \xi,\epsilon \right ]$) is used to represent the error and rotation angle, which is shown as (\ref{e:att_quat1}) and (\ref{e:att_quat2}).
\begin{equation}
\label{e:att_k}
\begin{array}{ll}
 \omega_{d} = -K\xi_{e}
 \end{array}
\end{equation}
\vspace{-0.75cm}

\begin{equation}
\label{e:att_quat1}
\begin{array}{ll}
 \theta =  2\cos^{-1} \left | \eta \right |
\end{array}
\end{equation} 
\vspace{-0.75cm}

\begin{equation}
\label{e:att_quat2}
 \begin{array}{ll}
 \displaystyle \xi_{e} =  sgn(\eta)\frac{\frac{\theta}{2}}{\sin\frac{\theta}{2}}\epsilon
 \end{array}
\end{equation}
\vspace{-0.25cm}

$ C $ is the angular velocity controller algorithm which is the main topic of the attitude and angular velocity loop. The loop-shaping technique will be used to design the controller with two steps. The first step is the frequency-domain model identification, and the second step is modal elimination and compensator design.
\vspace{0.25cm}
\subsubsection{System Identification}

\begin{figure}[h]
    \begin{center}
        \vspace{-0.25cm}
        {\includegraphics[width=1\columnwidth]{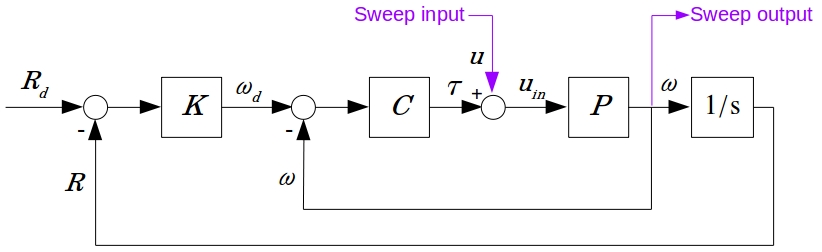}} 
    \end{center}
    \vspace{-0.25cm}
    \caption{\label{fig:sysident_in_out}the location of input and output signals}
        \vspace{-0.25cm}
\end{figure}

$ {}\vspace{0.25cm}\\ $
\indent The angular velocity controller is designed through the loop-shaping method which is a frequency domain method. To do so, we need to obtain the aircraft model in the frequency domain. (\ref{e:att_quat2}) presents the first principle model for the aircraft rate dynamics, however it fails to capture the detailed features, such as flexible modes. Therefore a sweep experiment is used to identify the frequency-domain model. The location of the injected signal and output data are shown in Fig. \ref{fig:sysident_in_out}. The sweep signal $ u $ is directly added to the output of the angular velocity controller $ \tau $. The angular velocity feedback $ \omega $ measured by onboard MEMS gyroscope is used as an output signal. The sampling rate of original gyroscope data is 1 kHz but is reduced to 250 Hz through a Butterworth filter to suppress the measurement noise. 250 Hz is also the controller update rate. 

\begin{figure}[h]
    \begin{center}
        \vspace{-0.25cm}
        {\includegraphics[width=1\columnwidth]{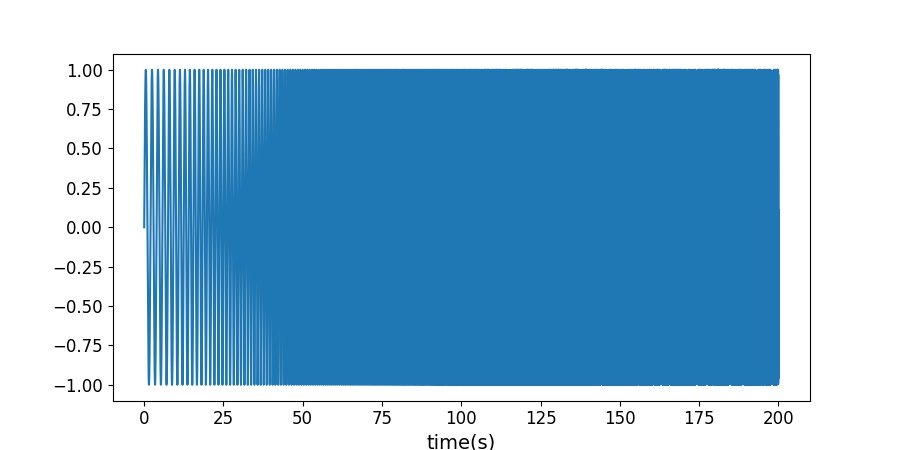}} 
    \end{center}
    \vspace{-0.25cm}
    \caption{\label{fig:sweep_sig}The exponential type chirp signal at time domain}
        \vspace{-0.25cm}
\end{figure}

\begin{figure}[h]
    \begin{center}
        \vspace{0.75cm}
        {\includegraphics[width=1.05\columnwidth]{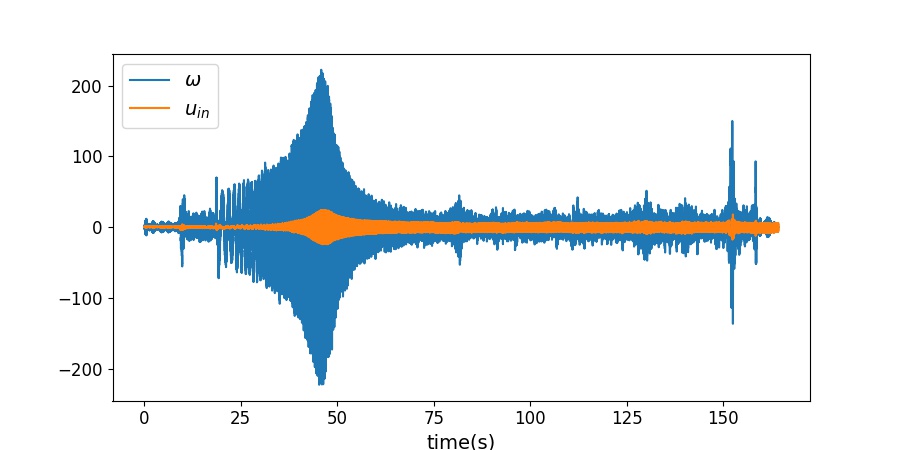}} 
    \end{center}
    \vspace{-0.25cm}
    \caption{\label{fig:sweep_input}The input and output data at time domain(pitch rate)}
        \vspace{-0.25cm}
\end{figure}

The exponential type chirp signal~\cite{novak2010nonlinear} is used as the sweep signal which is given as the following equations:
\begin{equation}
\label{e:chirp1}
\begin{array}{ll}
 \displaystyle k = \left(\frac{f_{1}}{f_{0}}\right)^\frac{1}{T}
\end{array}
\end{equation} 
\vspace{-0.5cm}

\begin{equation}
\label{e:chirp2}
\begin{array}{ll}
 \displaystyle \varphi(t) = 2\pi f_{0}\frac{k^{t}-1}{\ln(k)}
\end{array}
\end{equation} 
\vspace{-0.5cm}

\begin{equation}
\label{e:chirp3}
 \begin{array}{ll}
 u(t) = A\sin[\varphi\left(t\right)]
 \end{array}
\end{equation}
\noindent where $u_{in}=\tau + u $ is the total input to the plant, $ [f_{0},\,f_{1}] $ is the range of frequencies that are of interest which is $ [1\,Hz,\,60\,Hz] $ for this part of work. $ T $ is the total time. $ t $ is the current time. A is the gain of the input, which is usually set to a suitable value in order to obtain a good signal to noise ratio. The time domain example data of sweep signal $ u $ is shown in Fig. \ref{fig:sweep_sig}. The time-domain input $ u_{in} $ and output $ \omega $ of pitch rate sweep experiment are shown in Fig. \ref{fig:sweep_input}.

\begin{figure}[h]
    \begin{center}
        \vspace{-0cm}
        {\includegraphics[width=1.08\columnwidth]{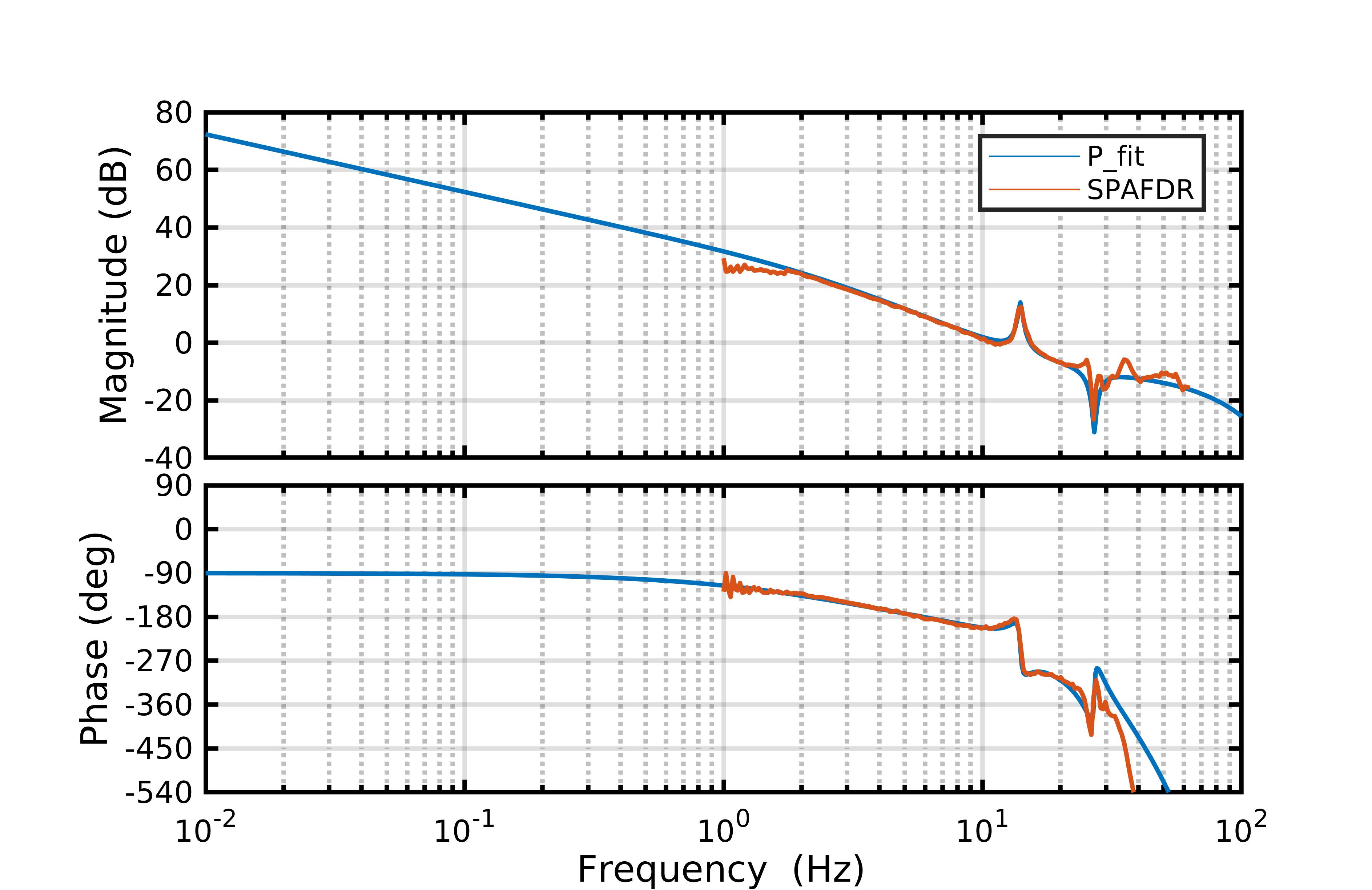}} 
    \end{center}
    \vspace{-0.25cm}
    \caption{\label{fig:freq_responce1}The actual frequency-domain model and fitted plant model}
        \vspace{-0.25cm}
\end{figure}

Based on the time domain data collected from the experiment, the Spectral Analysis with Frequency-Dependent Resolution (SPAFDR) is applied to process the experimental data and extract the system gain and phase delay at different frequencies. The details can be found in~\cite{ljung2003version}. Here we show the spectrum analysis data and fitted model in Fig. \ref{fig:freq_responce1}. The blue line stands for the model fitted in the continuous frequency domain while the orange represents the frequency response data extracted by SPAFDR.

The fitted model consists of multiple components as sheen in (\ref{e:plant_c1}). First, for the Pixhawk 4 flight stack, a second-order Butterworth low-pass filter $ P_{lf} $ at $ 69\,Hz $ has been used to isolate the disturbance coming from rotors whose frequency is between $ 75\,Hz $ and $ 90\,Hz $. This low pass filter is lumped into the aircraft system when identifying the model and therefore is used as a component in the fitted model (i.e., (\ref{e:plant_c3})). Then, we use a second order transfer function $ P_{D_y} $ (\ref{e:plant_c4}) which consists of an integrator, two zeros, and one pole to approximate the system main dynamics. Furthermore, noticing that the frequency response data has a peak and off peak at $ 14 $ and $ 27\, Hz $ respectively, we also include a peak (\ref{e:plant_c5}) and offpeak component (\ref{e:plant_c6}) in the model. Finally, a pure delay component (\ref{e:plant_c7}) is used to model the system delay. 

\begin{equation}
\label{e:plant_c1}
 \begin{array}{ll}
 \displaystyle P = P_{lf}P_{Dy}P_{peak}P_{offpeak}P_{delay}
 \end{array}
\end{equation}
\vspace{-0.75cm}

\begin{equation}
\label{e:plant_c3}
 \begin{array}{ll}
 \displaystyle P_{lf} = \frac{1}{1+0.00321s+0.00000531s^{2}}
 \end{array}
\end{equation}
\vspace{-0.25cm}

\begin{equation}
\label{e:plant_c4}
 \begin{array}{ll}
 \displaystyle P_{Dy} = \frac{260+3.764s+0.01362s^{2}}{1+0.0637s} \frac{1}{s} 
 \end{array}
\end{equation}
\vspace{-0.25cm}

\begin{equation}
\label{e:plant_c5}
 \begin{array}{ll}
 \displaystyle P_{peak} = \frac{1+0.00239s+0.000129s^{2}}{1+0.000341s+0.000129s^{2}}
 \end{array}
\end{equation}
\vspace{-0.25cm}

\begin{equation}
\label{e:plant_c6}
 \begin{array}{ll}
 \displaystyle P_{offpeak} = \frac{1+0.000118s+0.0000348s^{2}}{1+0.0013s+0.0000348s^{2}}
 \end{array}
\end{equation}
\vspace{-0.25cm}

\begin{equation}
\label{e:plant_c7}
 \begin{array}{ll}
 \displaystyle P_{delay} = e^{-0.021s}
 \end{array}
\end{equation}
\vspace{-0.25cm}

\vspace{0.25cm}
\subsubsection{Notch Filter Design}
\begin{figure}[ht]
    \begin{center}
        \vspace{-0.25cm}
        {\includegraphics[width=1.05\columnwidth]{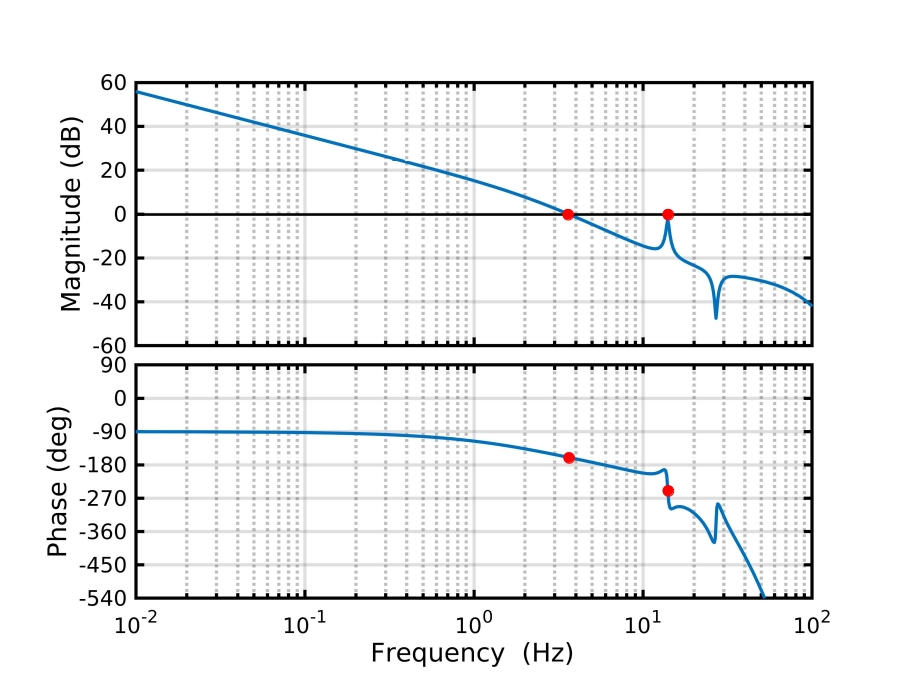}} 
    \end{center}
    \vspace{-0.25cm}
    \caption{\label{fig:vib_14hz}The influence of peak near 14 Hz}
        \vspace{-0.25cm}
\end{figure}

$ {}\vspace{0.25cm}\\ $
\indent Through the frequency-domain model identification, there exists a strong peak near the $ 14\,Hz $ in the spectrum where the phase near $ 14\,Hz $ is less than $ -180\,^\circ $. This pitch represents a resonance mode of the UAV structure, and if handled properly, the pitch rate loop will vibrate considerably or even become unstable if the magnitude is above $ 0\,dB $ at this resonant frequency, as shown in Fig. \ref{fig:vib_14hz}. Indeed, we tuning the controller, we found that the four motor brackets will vibrate divergently due to this resonance mode. This circumstance will limit the bandwidth of pitch control. For this aircraft, the bandwidth will be limited to less than $ 4\,Hz $, as implied by Fig. \ref{fig:vib_14hz}. When further increasing the controller bandwidth (controller gain), the resonance mode will cross $ 0\,dB $ and the system will become unstable. Enhancing the structure stiffness can rise the location of the resonance frequency, thus allowing further increment in the controller bandwidth. However, increasing the stiffness usually leads to adding more materials, thus increasing the UAV weight. To solve this problem with the lowest cost, the notch filter is used to eliminate this peak in the spectrum.

\begin{figure}[h]
    \begin{center}
        \vspace{-0.4cm}
        {\includegraphics[width=1.05\columnwidth]{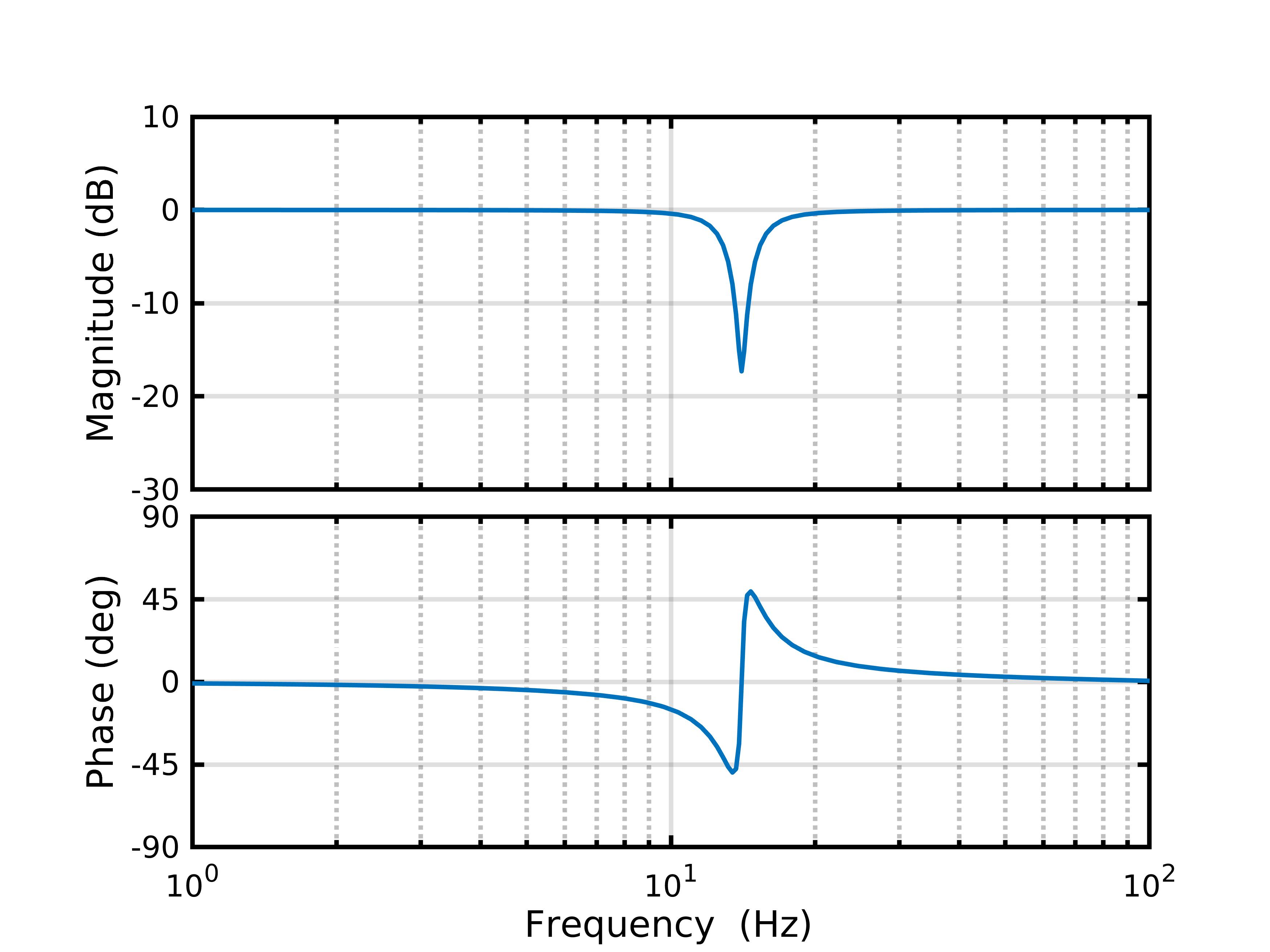}} 
    \end{center}
    \vspace{-0.25cm}
    \caption{\label{fig:notch}The frequency response of notch filter}
        \vspace{-0.25cm}
\end{figure}

\begin{figure}[h]
    \begin{center}
        \vspace{-0.5cm}
        {\includegraphics[width=1.05\columnwidth]{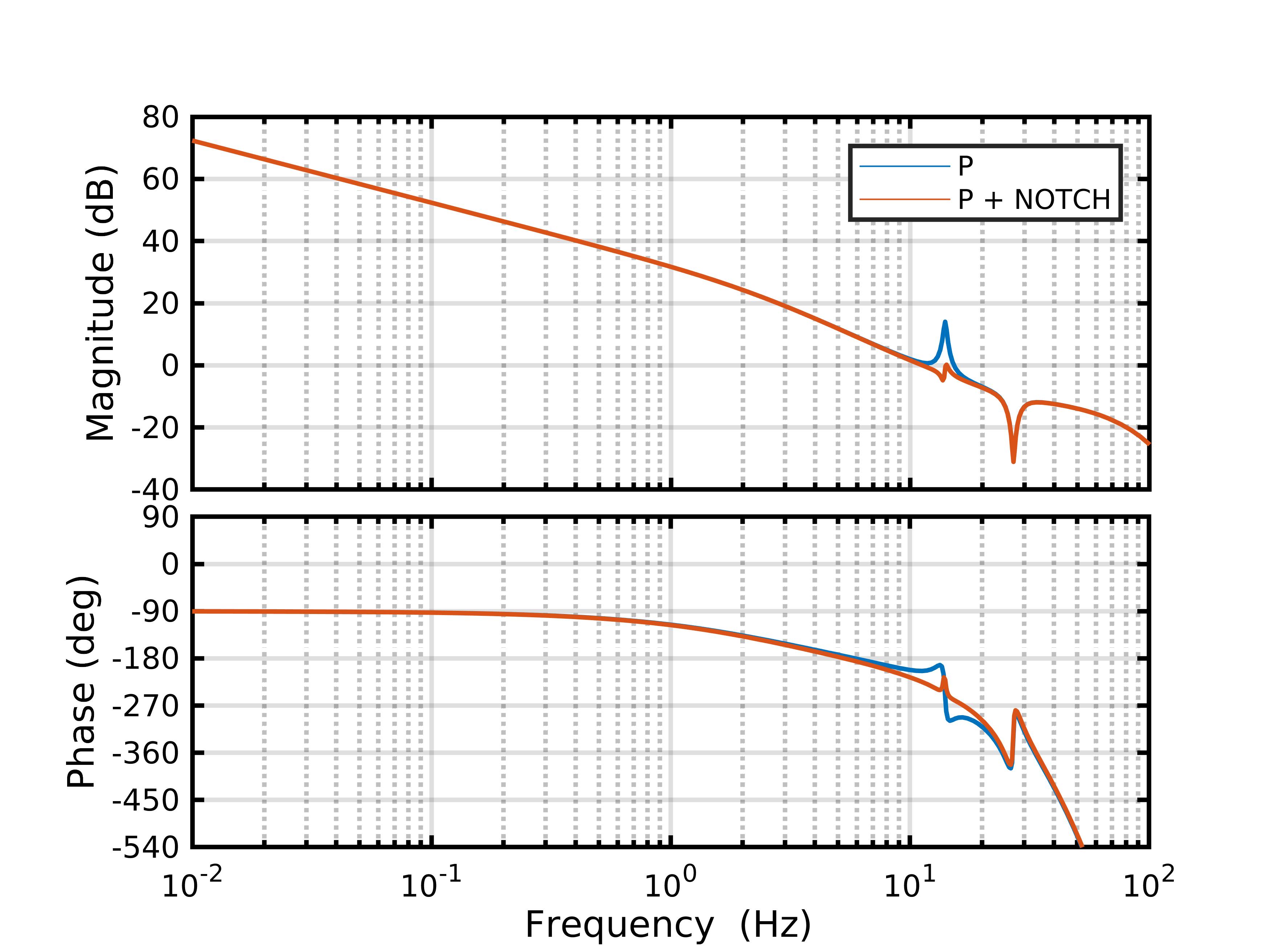}} 
    \end{center}
    \vspace{-0.25cm}
    \caption{\label{fig:comp_notch}The frequency response of notch filter}
        \vspace{-0.25cm}
\end{figure}

A notch filter is a type of the band-stop filter but with very sharp pass-band. (\ref{e:notch_tf1}) and (\ref{e:notch_tf2}) is the transfer function of the notch filter used in the pitch angular velocity loop:

\begin{equation}
\label{e:notch_tf1}
 \begin{array}{ll}
 \displaystyle N(s) = \frac{as^{2}+cs+1}{as^{2}+bs+1}
 \end{array}
\end{equation}
\vspace{-0.5cm}

\begin{equation}
\label{e:notch_tf2}
 \begin{array}{ll}
 \displaystyle a = \frac{1}{\omega_{0}^{2}} \quad b=\frac{k_{1}}{\omega_{0}} \quad c=\frac{k_{2}}{\omega_{0}}
 \end{array}
\end{equation}

\noindent where $ k_{1} $ and $ k_{2} $ is the width and depth of notch shape (Fig. \ref{fig:notch}), $ \omega_{0} $ is the central frequency. Fig. \ref{fig:notch} shows the frequency response of notch filter. The notch filter will lead to the reduction of phase margin. Therefore the $ k_{1} $ and $ k_{2} $ should be set to suitable values. Based on the frequency domain model from the last section, the notch filter will lead to a $ 5\,^\circ $ phase reduction at $ 7\,Hz $ which is the estimated controller bandwidth with good robustness margin. The influence of the notch filter when added to the pitch rate loop is shown in Fig. \ref{fig:comp_notch}.

\vspace{0.25cm}
\subsubsection{PID compensator design}
\begin{figure}[h]
    \begin{center}
        {\includegraphics[width=0.9\columnwidth]{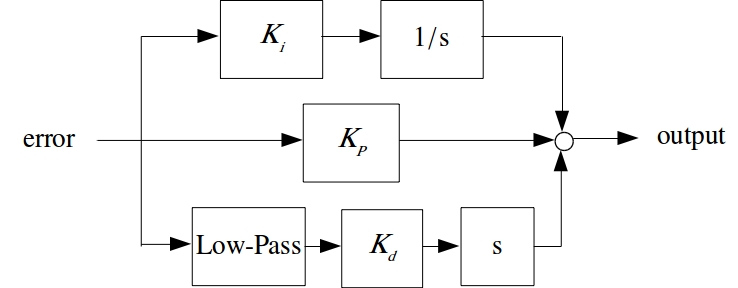}} 
    \end{center}
    \vspace{-0.25cm}
    \caption{\label{fig:pid_struc}The structure of PID compensator}
        \vspace{-0.25cm}
\end{figure}

$ {}\vspace{0.25cm}\\ $
\indent The main concept of the loop shaping technical is designing the open-loop transfer function to optimize the performance of closed-loop frequency response. Based on the unit feedback, the closed-loop transfer function $ T $ can be denoted as $  T = PC/(1+PC) $, where the $ P $ and $ C $ can be found in Fig. \ref{fig:atti_and_rate}. For the low-frequency section, $ |PC| $ should be high enough to resist the low-frequency disturbance(e.g., static disturbances like the difference between motors). For the mid-frequency section, the $ PC $ should be designed as approximately a first-order system to increase robustness. For the high-frequency section, $  |PC| $ should be low enough to make sure the closed-loop stability. There are several primary components that can be used to shape the loop transfer function $ PC $, such as lead, lag, roll off, PID compensator. For this paper, the PID compensator is used to design the loop shaping controller. Fig. \ref{fig:pid_struc} shows the structure of the PID compensator.

\begin{figure}[h]
    \begin{center}
        \vspace{-0.25cm}
        {\includegraphics[width=1.05\columnwidth]{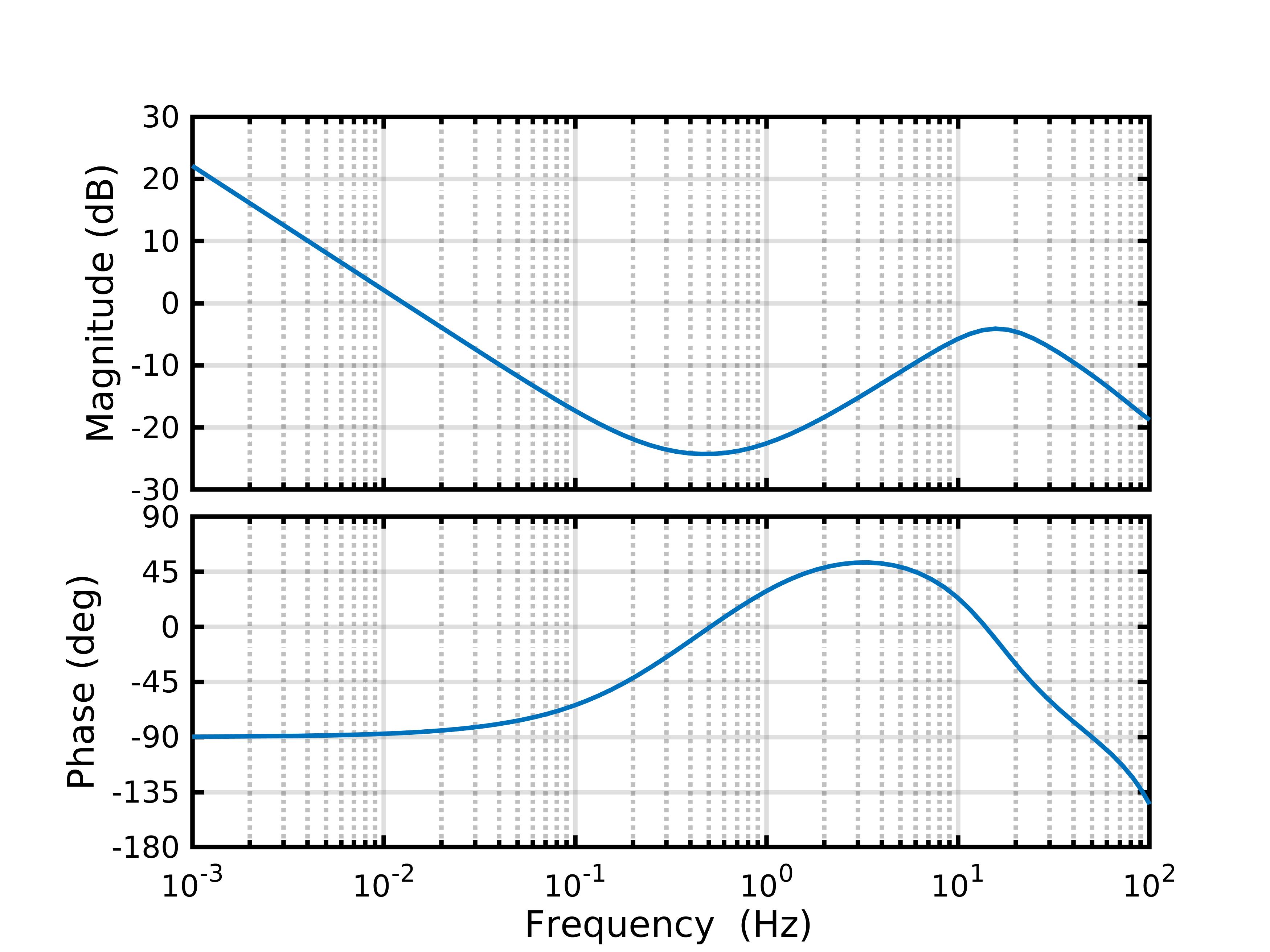}} 
    \end{center}
    \vspace{-0.25cm}
    \caption{\label{fig:pid_comp}The frequency response of PID compensator}
        \vspace{0.25cm}
\end{figure}

Fig. \ref{fig:pid_comp} shows a general frequency response of the PID compensator. The integral action boosts the gain of the loop transfer function at low frequency. The differentiator with low-pass filter adds extra phase lead at the cross over frequency such that the phase margin is increased. The low-pass filter used here is a second-order Butterworth filter which decreases the high-frequency gain to eliminate the influence of high-frequency measurement noises due to the effect of from rotors or medal.

\begin{figure}[h]
    \begin{center}
        \vspace{0.5cm}
        {\includegraphics[width=1.05\columnwidth]{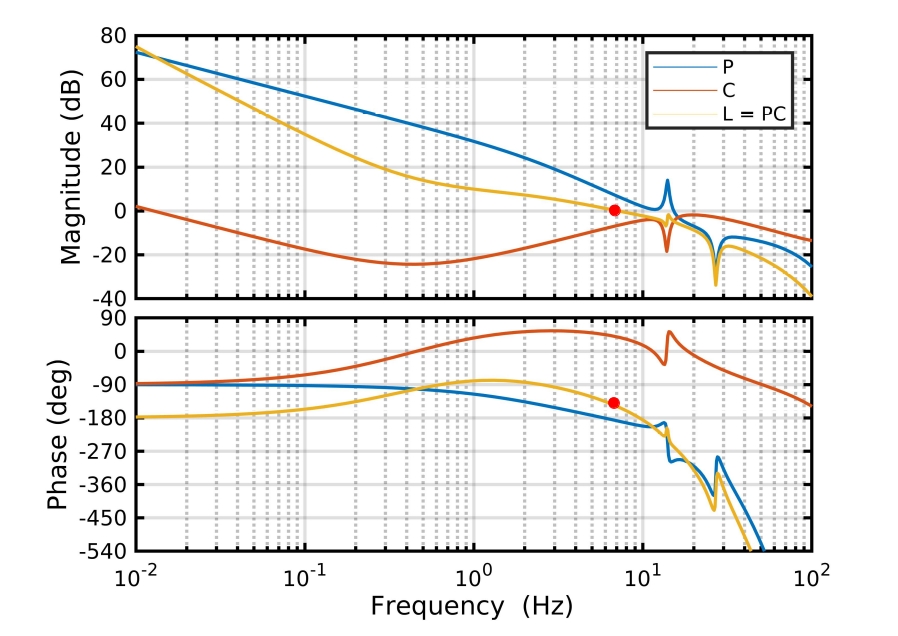}} 
    \end{center}
    \vspace{-0.5cm}
    \caption{\label{fig:bode_PCL}The open-loop frequency response of pitch angular velocity loop}
        \vspace{-0.75cm}
\end{figure}

For our prototype, $ K_{i} $, $ K_{p} $ and $ K_{d} $ of the PID compensator are set to 0.1, 0.09 and 0.01 respectively, the corner frequency of the low pass filter in the differential action is set to $ 18\,Hz $. The frequency response of the designed controller is shown in Fig. \ref{fig:bode_PCL}, where the C is the PID compensator with the notch filter. The bandwidth of pitch angular velocity is $ 6.8\, Hz $ which is $ 70\,\% $ more than the maximum bandwidth without notch filter. The phase margin is $ 44\,^\circ $ which is in the desired range for typical aircraft systems. Between $ 0.6\,Hz $ to $ 14\,Hz $, the slope of magnitude is $ -19\, dB/dec $ which is the robust frequency section of pitch angular rate. 

\subsection{Altitude Controller}
\label{sec:altitude_controller}
To simplify the altitude controller design, the aircraft is always in coordinated flight which refers to a flight state with no side-slip~\cite{ritz2017global}. Coordinated flight means the the the aerodynamic lift is only the function of airspeed and angle of attack(AOA) when the environment and aircraft configure parameter are steady. The belows will describe the altitude controller design based on the coordinated flight assumption.

\vspace{0.25cm}
\subsubsection{Aerodynamic Force Model}
$ {}\vspace{0.25cm}\\ $
\indent In our previous work, the full aerodynamic parameters for another tail-sitter UAV which has similar fly-wing configuration and dimension parameters with the current one was captured in the wind tunnel experiment~\cite{wang2017design}. Here we use these data to model current aircraft's aerodynamic force as Fig. \ref{fig:aero} shown. Based on the coordinated flight assumption, the aerodynamic lift $ L $ and drag $ D $ can be calculated as (\ref{e:aero_5}) and (\ref{e:aero_6}):

\begin{figure}[h]
    \begin{center}
        \vspace{0.25cm}
        {\includegraphics[width=1.05\columnwidth]{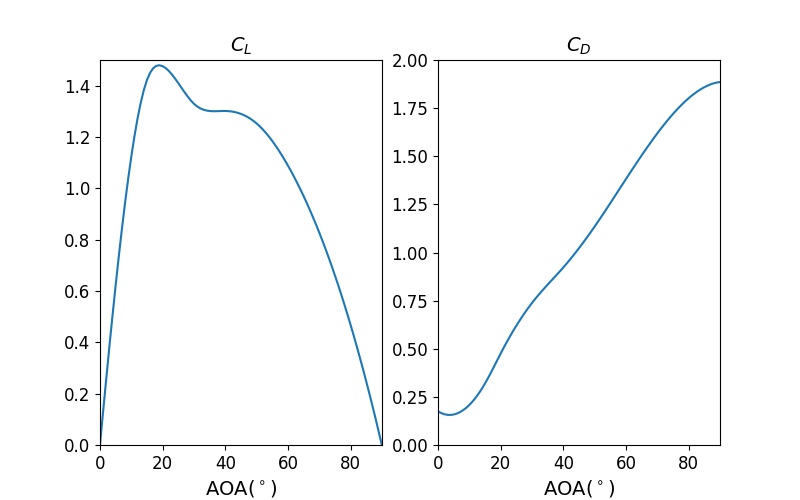}} 
    \end{center}
    \vspace{-0.25cm}
    \caption{\label{fig:aero}The aerodynamic model}
    \vspace{-0.5cm}
\end{figure}

\begin{equation}
\label{e:aero_5}
 \begin{array}{ll}
 \displaystyle L = \frac{1}{2}\rho V^{2} S C_{L}(\alpha,V)
 \end{array}
\end{equation}
\vspace{-0.65cm}

\begin{equation}
\label{e:aero_6}
 \begin{array}{ll}
 \displaystyle D = \frac{1}{2}\rho V^{2} S C_{D}(\alpha,V)
 \end{array}
\end{equation}
\noindent where $ \rho $ is the air density, $ \alpha $ is the angle of attack. The ground speed is used as airspeed $ V $ assuming the wind speed is small enough. $ C_{L} $ and $ C_{D} $ is the aerodynamic lift coefficient and drag coefficient respectively.

\vspace{0.25cm}
\subsubsection{Controller Design}
\begin{figure}[h]
    \begin{center}
        {\includegraphics[width=1\columnwidth]{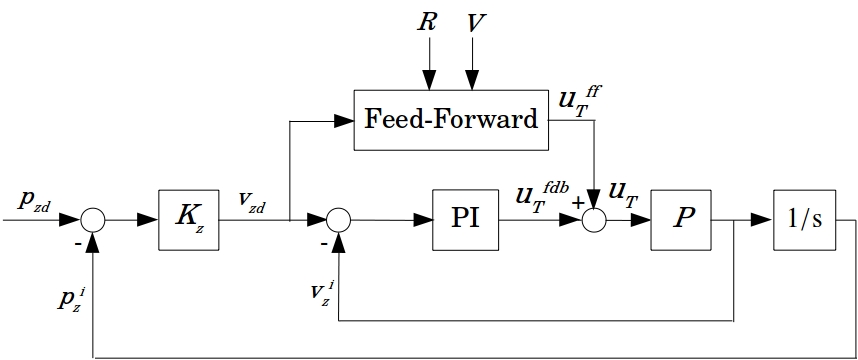}} 
    \end{center}
    \vspace{-0.25cm}
    \caption{\label{fig:altitude-struct}The detailed structure of the altitude controller}
    \vspace{-0.25cm}
\end{figure}

$ {}\vspace{0.25cm}\\ $
\indent The structure of altitude controller is shown as Fig. \ref{fig:altitude-struct}. There is a feedforward-feedback parallel structure to follow the target vertical velocity $ v_{zd} $. The feedforward controller is a proportional controller, which computes the desired acceleration $ a_{zd} $ from the desired velocity $ v_{zd} $ by multiplying it by a proportional constant (i.e., (\ref{e:thrust1})). The desired acceleration in $ z^{i} $ direction is then used to determine the propeller thrust $ T $ through solving the (\ref{e:thrust2}) which is the combination of (\ref{e:plant12}), (\ref{e:aero_2}), and (\ref{e:aero_3}). Finally, the required thrust $ T $ is used to compute the controller commands $ u_T^{ff} $ as shown in (\ref{e:thrust3}) where the constant k is the ratio between the actual thrust in hovering and the corresponding control commands (\ref{e:thrust4}).

\begin{equation}
\label{e:thrust1}
 \begin{array}{ll}
 a_{zd}=K_{ff}v_{zd}
 \end{array}
\end{equation}
\vspace{-0.75cm}

\begin{equation}
\label{e:thrust2}
\begin{array}{ll}
 a_{zd}=mg+e_{3}\left(R_{v}^{i}\begin{bmatrix}-D\\ 0\\ -L
 \end{bmatrix} + R_{b}^{i}
 \begin{bmatrix}T\\ 0\\ 0
 \end{bmatrix}\right)
\end{array}
\end{equation}
\vspace{-0.25cm}


\begin{equation}
\label{e:thrust3}
 \begin{array}{ll}
 u_{T}^{ff}=kT
 \end{array}
\end{equation}
\vspace{-0.75cm}

\begin{equation}
\label{e:thrust4}
 \begin{array}{ll}
 \displaystyle k = \frac{T_{h}}{mg}
 \end{array}
\end{equation}

The model in the feedforward controller could suffer modeling mismatch. Therefore, a PI feedback controller is also used in parallel to the feedforward controller to correct the output of the altitude controller.

\section{Experimental Verification}
\label{sec:Experiments}

In this section, flight tests are provided to verify the proposed control system.

\subsection{Hover Test}
\begin{figure}[h]
    \begin{center}
        \vspace{-0.25cm}
        {\includegraphics[width=1\columnwidth]{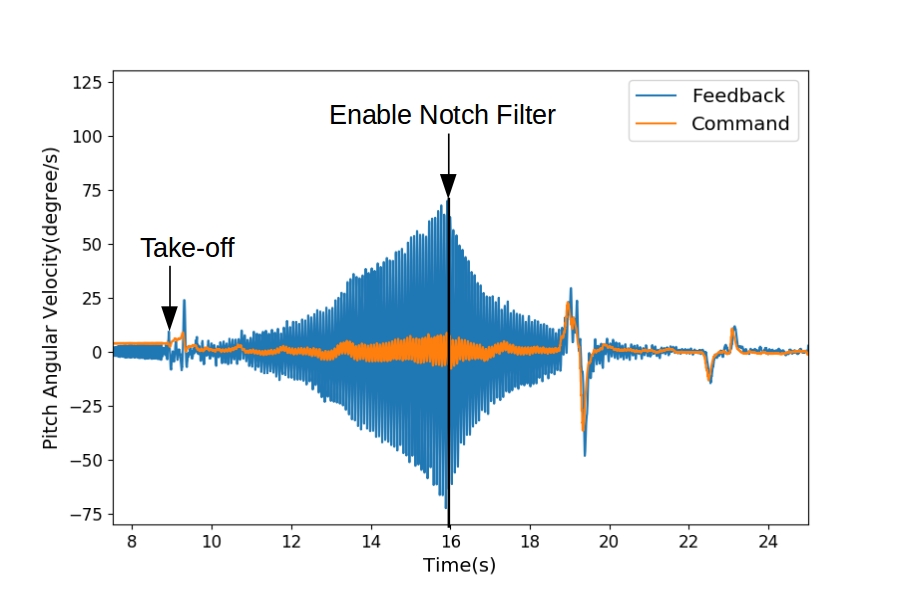}} 
    \end{center}
    \vspace{-0.25cm}
    \caption{\label{fig:notch_test}The notch filter test when hovering}
        \vspace{-0.25cm}
\end{figure}

\begin{figure}[h]
    \begin{center}
        \vspace{-0.25cm}
        {\includegraphics[width=1\columnwidth]{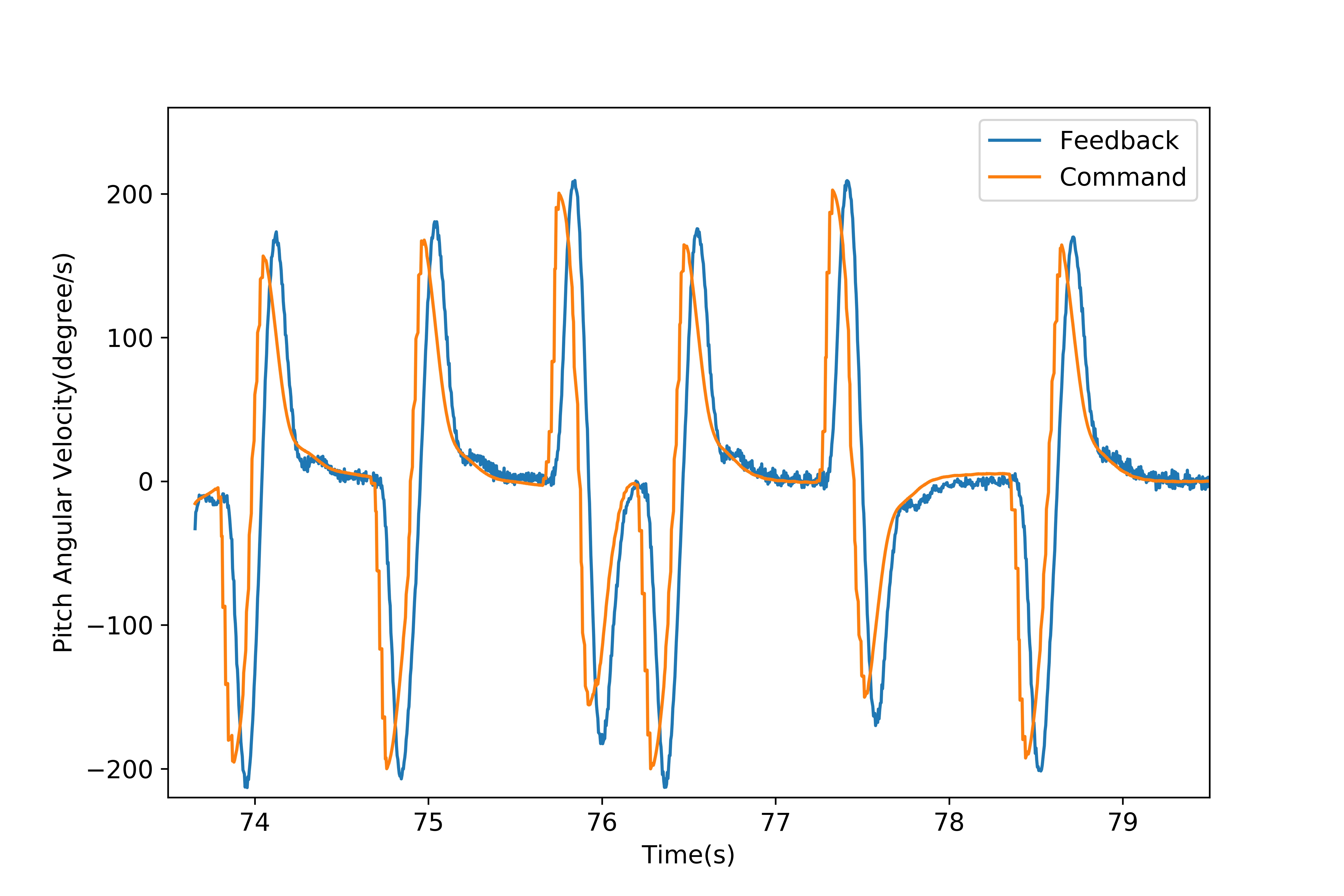}} 
    \end{center}
    \vspace{-0.5cm}
    \caption{\label{fig:comp_pr}The pitch angular velocity response after loop-shaping}
        \vspace{-0.5cm}
\end{figure}

To verify the effectiveness of the notch filter, the notch filter is disabled at the beginning. It is clearly shown in Fig. \ref{fig:notch_test} that the pitch loop is vibrating after take-off, and the vibration amplitude becomes greater and greater which means the system is unstable without the notch filter. After the notch filter is enabled, the divergent oscillation turns into convergence rapidly. Test video can be seen in the https://youtu.be/ejllxRIaBQ0. This test verifies the analysis in Fig. \ref{fig:vib_14hz} and Fig. \ref{fig:comp_notch}. Fig. \ref{fig:comp_pr} shows the pitch rate time-domain response after loop-shaping when tracking the fast manual command. It is not hard to observe that the loop-shaping controller exhibits none overshoot and fast-tracking.

\subsection{Forward Flight Test}

\begin{figure}[h]
    \begin{center}
    \vspace{-0.25cm}
        {\includegraphics[width=1.03\columnwidth]{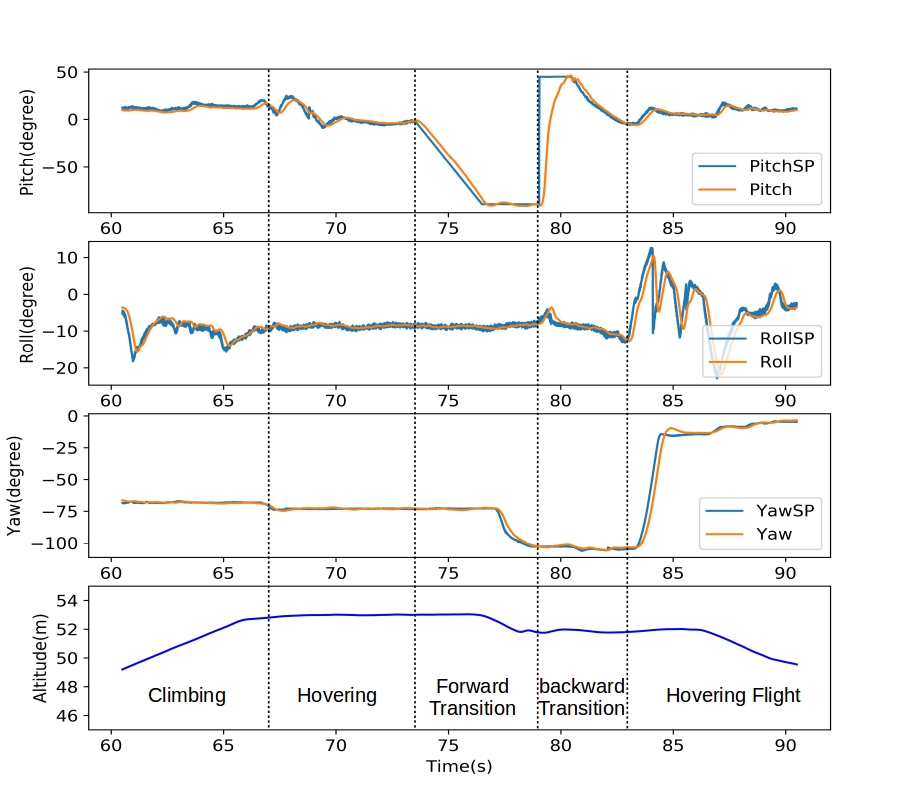}} 
    \end{center}
    \vspace{-0.5cm}
    \caption{\label{fig:forward_test}Attitude and altitude response at forward flight test}
        \vspace{-0.25cm}
\end{figure}

To verify the robustness of the attitude controller, a forward accelerating and decelerating test is conducted. Fig. \ref{fig:forward_test} shows the whole attitude response at whole test. At the start, the aircraft is climbing at low speed. When the desired altitude is achieved at 66 s, the aircraft hover for some time. Then after 73 s, linear forward tilt command is given to the pitch attitude controller to let the aircraft accelerate. When the pitch angle comes to $85\,{}^\circ $, the aircraft will stay at a current pitch angle to keep forward flight for some time. It is clear that all the controllers track their commands quickly and with almost no overshoot during this process. After 79 s, a backward step command is given to the pitch channel to let the aircraft come back to hovering. The step response in pitch direction shows that the attitude loop behaves as a first order system. Fig. \ref{fig:forward_test} also shows the altitude change during the forward accelerating and decelerating flight. It is obvious that the altitude error during the whole flight is less than 2 m which verify the altitude controller design.

\begin{figure}[h]
    \begin{center}
    \vspace{-0.25cm}
        {\includegraphics[width=1.075\columnwidth]{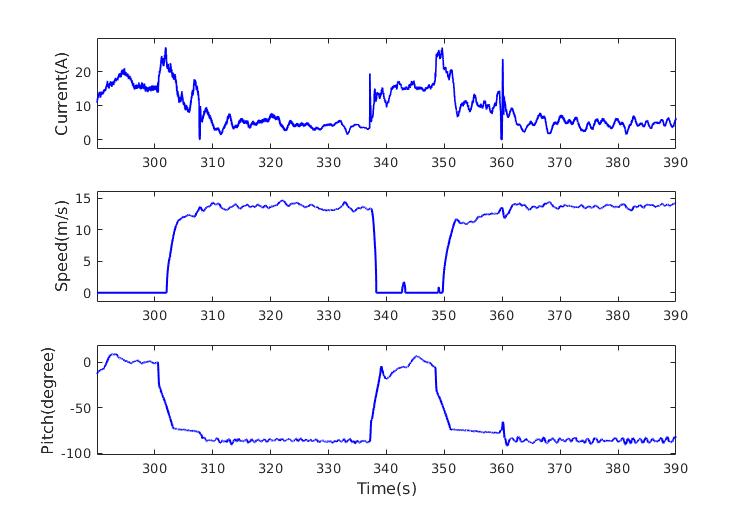}} 
    \end{center}
    \vspace{-0.5cm}
    \caption{\label{fig:eff_test}Current, speed and pitch of the outdoor flight test}
        \vspace{-0.25cm}
\end{figure}

Besides the transition test, another flight test in a wider open area, which is $ 500m\times 600m $, is conducted to verify the UAV performance in level flight. Fig. \ref{fig:eff_test} shows the current, speed and pitch angle during flight. In the sections $ [290\,s,\,300\,s] $ and $ [340\,s,\,348\,s] $, the aircraft is hovering with the current near to $ 20\,A $. In the sections $ [310\,s,\,336\,s] $ and $ [360\,s,\,390\,s] $, the aircraft is flying as $ 13.67\, m/s $ with the average current $ 5\, A $ which is around a quarter of the current in hovering. This test result proves the flight efficiency of this carbon-fiber structured tail-sitter aircraft.

\section{Conclusion}
\label{sec:conclusion}

In this study, We have proposed a unified control system that can work in the high angles of attack. The control system mainly consists of an attitude controller and an altitude controller. To optimize the attitude controller, the frequency domain model is identified through the frequency sweep experiment in which the continuous exponential chirp signal is used as sweep input. Based on the frequency domain model of the aircraft, a notch filter is designed to decrease the influence of modes near to $ 14\, Hz $. Then a PID compensator is used, and its parameters are tuned via loop-shaping method. The altitude controller consists of a feedforward controller and feedback controller. The thrust command is calculated based on the current speed and attitude. The hover and forward flight tests prove the effectiveness of the above designs and methods.

\bibliography{paper} 
\end{document}